\documentclass[aps,prb,twocolumn,showpacs,groupedaddress]{revtex4}
\usepackage{graphicx}
\DeclareGraphicsExtensions{.jpg, .pdf, .png}

\begin{document}

\title{Electrical resistivity of the Ti$_4$O$_7$ Magneli phase under high
pressure}

\author{C. Acha}
\thanks{Also fellow of CONICET of Argentina}
\email{acha@df.uba.ar}
\author{M. Monteverde}
\thanks{Scholarship of CONICET of Argentina}
\affiliation{Laboratorio de Bajas Temperaturas, Departamento de F\'{\i}sica,
FCEyN, Universidad Nacional de Buenos Aires, Pabell\'on I, Ciudad
Universitaria, 1428 Buenos Aires, Argentina}
\author{M. N\'u\~nez-Regueiro}
\affiliation{CRTBT, CNRS, BP166, 38042 Grenoble Cedex 09, France}
\author{A. Kuhn}
\affiliation{Facultad de Ciencias Experimentales y T\'ecnicas, Universidad San
Pablo, 28668 Boadilla del Monte, Espa\~na}
\author{M. A. Alario Franco}
\affiliation{Facultad de Ciencias Qu\'{\i}micas, Universidad Complutense,
28040 Madrid, Espa\~na}



\date{\today}

\begin{abstract}

We have measured resistivity as a function of
temperature and pressure of Ti$_4$O$_7$ twinned crystals using
different contact configurations. Pressures over 4 kbar depress
the localization of bipolarons and allow the study of the
electrical conduction of the bipolaronic phase down to low
temperatures. For pressures $P>40$ kbar the bipolaron formation
transition is suppressed and a nearly pressure independent
behavior is obtained for the resistivity. We observed an
anisotropic conduction. When current is injected parallel to the
principal axis, a metallic conduction with interacting carrier
effects is predominant. A superconducting state was not obtained
down to 1.2 K, although evidences of the proximity of a quantum
critical point were noticed. While when current is injected
non-parallel to the crystal's principal axis, we obtained a
logarithmic divergence of the resistivity at low temperatures. For
this case, our results for the high pressure regime can be
interpreted in the framework of interacting carriers (polarons or
bipolarons) scattered by Two Level Systems.
\end{abstract}

\maketitle




\section{INTRODUCTION}

The study of electronic properties in metals which their charge
carriers could be associated with bipolarons has recently raised
some interesting discussions in the quest to determine the
microscopic origin of high T$_c$ superconductivity
\cite{Chakra98,Alex99,Chakra99}. Alexandrov and Mott \cite{Alex94}
had claimed that many of the experimental properties of high
temperature superconductors can be explained considering a Bose
condensation of their bipolaronic charge carriers. Chakraverty et
al. \cite{Chakra98} argue that the condensation temperature for
this type of bosonic superconductors can not be much higher than
10 K. The possibility of a superconductivity related to bipolarons
was previously studied theoretically by several authors
\cite{Alex81,Iguchi88,Micnas90}, but up to now there is not a
clear experimental proof of the existence of a superconducting
state based on a Bose condensation of bipolarons.

On the other hand, characteristic transport properties of a
bipolaronic metal are not well established experimentally.
Theoretical studies \cite{Alex97} asserted a logarithmic
divergence of the resistivity at low temperatures, in the case of
bipolaronic carriers scattered by shallow potential wells related
to disorder. This type of divergence was particularly observed in
the anisotropic low-temperature normal-state resistivity of
underdoped La$_{2-x}$Sr$_x$CuO$_4$ single crystals, where
superconductivity was suppressed with a 60 T pulsed magnetic field
\cite{Ando95,Ando96}.

\begin{figure}
\includegraphics[width=3in]{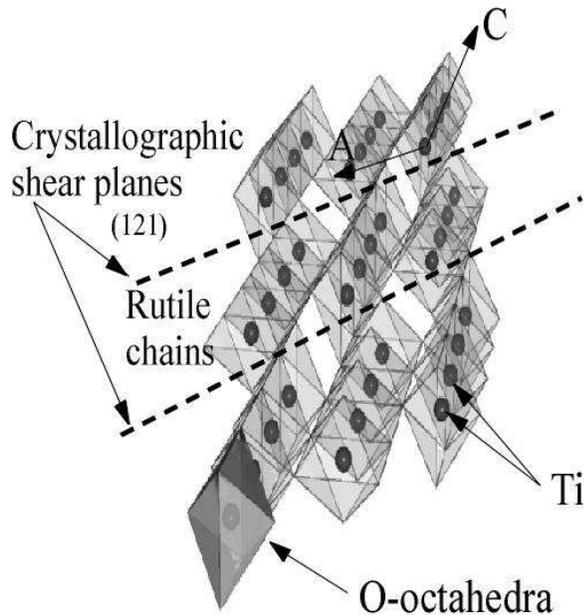}
\vspace{5mm}
\caption{Crystal structure of Ti$_4$O$_7$. This structure is
formed by successive rutile blocks (TiO$_2$), infinite in 2
dimensions and of 4 octahedra  width along the $C$ axis, bounded
by the crystallographic shear planes, corresponding to the (121)
planes referred to the rutile structure.}

\vspace{5mm} \label{estructura}
\end{figure}

In order to contribute to this debate and to gain some insight in
transport properties related to (bi)polaron diffusion,  we
considered the study of a material with a conduction related to
these quasi-particles. The low temperature electrical conduction
of the Ti$_4$O$_7$ Magneli phase can be assign to bipolaron
diffusion as was revealed by electrical resistivity, magnetic
susceptibility, specific heat, ESR and crystal structure studies
\cite{Schlenker74,Chakra76,Lakkis76,Schlenker79,Schlenker80,LePage84}.
This compound shows metal-non metal transitions (3 phases and 2
resistive transitions, with related specific heat jumps) which are
probably associated to the fact that it has one 3d electron per
two cation sites, and two valence states for the Ti (3+ and 4+).
The crystal structure of Ti$_4$O$_7$, shown in Fig.
\ref{estructura} , is built up by infinite (TiO$_6$) octahedra in
two dimensions, forming a rutile slab, having a finite width of n
octahedra along the $C$ crystallographic direction. The slabs are
delimited by the shear planes, referred as the (121) planes of the
rutile structure. The octahedra share corners, faces or borders,
determining 4 different crystallographic sites for the Ti.

For the low temperature phase (LTP) ($T \leq  T_{bl}$, with
$T_{bl}$ $\simeq$ 130 K (on cooling) or 140 K (on heating)  being
the bipolaron localization temperature) an insulator behavior is
found, where the Ti$^{3+}$ are well localized in chains of
Ti$^{3+}$-Ti$^{3+}$ pairs, forming a non magnetic bond and
creating a local distortion of the lattice. This pair, stabilized
by the lattice deformation is known as a bipolaron
\cite{Anderson75}. For the intermediate temperature phase (ITP)
($T_{bl}< T \leq T_{bf}$, with $T_{bf} \simeq$ 150 K, being the
bipolaron formation temperature) bipolarons become thermally
diffused and there is a correlative increase of the electrical
conduction. In the high temperature phase (HTP) ($T>T_{bf}$), the
3d electrons become delocalized; a sudden increase of the magnetic
susceptibility (Pauli paramagnetism) is observed while the
conduction becomes metallic\cite{Schlenker80}.

It was tempting to extend towards lower temperatures the existence
window of the ITP. On one hand, to favor the conditions for a
possible bipolaron condensation that would yield a superconducting
state. On the other hand, the study of transport properties of
such a material can reveal new features, besides the ones that
were predicted theoretically \cite{Alex97} but not confirmed
experimentally.

V impurities were previously introduced for this purpose
\cite{Schlenker79,Schlenker80,Schlenker85}. They create local
structural distortions which generate long range interactions that
stabilize the ITP. Indeed, the Ti$^{3+}$ - Ti$^{3+}$ distance is
lower in the V doped sample and the temperature of the
metal-insulator transition is depressed. Unfortunately this
induces at the same time the increase of the background disorder
which produces a glassy conduction.

By applying an external pressure, structural distortions can be
generated in order to extend the IT phase, but without the
disadvantage of introducing impurities. In accordance to this, it
was shown that the formation and localization of bipolarons can be
both depressed applying pressures up to 40 kbar \cite{Acha96}. The
rate of depression of the localization temperature is higher than
the formation one. The temperature width for bipolaron diffusion
is then enhanced and their transport properties can be studied
down to low temperatures.

In this paper we report DC  resistivity measurements (R) as a
function of temperature ($T$) and pressure ($P$) of Ti$_4$O$_7$
twinned crystals, for 4 K$<T<$300 K and $P$ up to 220 kbar. An
anisotropic electrical transport was observed and analyzed for the
whole pressure regime. Pressure depress the localization and the
formation transitions yielding to a conducting regime down to low
temperatures. A metallic behavior with interacting carriers
effects and signs of the proximity of a quantum critical point is
obtained when current is applied parallel to the principal axis of
crystals, while a logarithmic divergence on the resistivity on
decreasing temperature is observed for currents applied with
components in a perpendicular direction.

\section{Experimental}

Ti$_4$O$_7$ powder was made by heating finely ground mixtures of
TiO$_2$ (rutile) and Ti metal in an evacuated silica tube at 1150
C during one week, as previously described \cite{SAndersson57}.
Small single crystals (0.1 $\times$ 0.1 $\times$ 0.3 to 0.6
mm$^3$) could be extracted from these powders. Bigger twinned
crystals (1 $\times$ 1 $\times$ 3 mm$^3$) were obtained by a
chemical transport reaction using Iodine as a transport agent. For
this, about 1g of freshly prepared Ti$_4$O$_7$ powder and 0.1g of
iodine were introduced in a silica tube, which was then sealed
under vacuum. The tube was placed in an horizontal furnace and
heated to 950 C during two weeks. The tube charge was held in the
hottest zone, and some well formed, blue-black crystals were
deposited in a colder region.

X-ray powder diffraction experiments were conducted to check the
purity of the obtained material. A Siemens Kristalloflex D-5000
diffractometer operating with CuK alpha radiation at a scan rate
of 0.1 degrees/min was used. Single crystal X-ray diffraction
experiments were performed on several as-grown specimens. The
crystals were stucked on a glass capillary and mounted in a
four-cycle Enraf-Nonius CAD4 diffractometer. The diffraction data
were collected using the 2$\theta\omega$ scan method with a scan
rate of 1 degree/min. at room temperature. The diffractometer was
operated at 40 kV/20 mA; a graphite - monochromatized MoK alpha
beam ($\lambda$ =0.71069) was used.

Samples were also characterized at ambient pressure by measuring
their ac susceptibility as a function of temperature. Using a
mutual inductance bridge, with an excitation signal frequency of
119 Hz, and 1 Oe of magnetic field amplitude, an overall
sensitivity of 10$^{-8}$ emu can be attained.

Resistivity under high pressure was measured using a hydrostatic
and a quasi-hydrostatic experimental setup depending on the
pressure range focused. For pressures bellow 10 kbar, a
hydrostatic piston cylinder self-clamped cell was used, while for
higher pressures, starting from 15 kbar and up to 220 kbar, we
applied the Bridgman configuration with sintered diamond anvils.
In the former case, a 50-50 mixture of kerosene and transformer
oil was used as the pressure transmitting medium and manganine as
a manometer. In the latter case, pyrophillite was used as a gasket
and steatite as the pressure medium that favors quasi-hydrostatic
conditions. Lead was used inside the cell as a manometer by
monitoring its superconducting transition temperature. The
pressure gradient was estimated from the width of this transition
and corresponds to a 5-10\% of the applied pressure, for pressures
$\leq$ 100 kbar, and saturates at a maximum value of 10 kbar for
higher pressures. We use a conventional 4 terminal DC technique to
measure resistivity under high pressure at different temperatures.
Electrical contacts were made using thin Pt wires, fixed to the
sample using silver paste for the hydrostatic cell, and solely by
pressure for the Bridgman setup.

\begin{figure} [htbp]
\begin{center}
\includegraphics[width=2.5in]{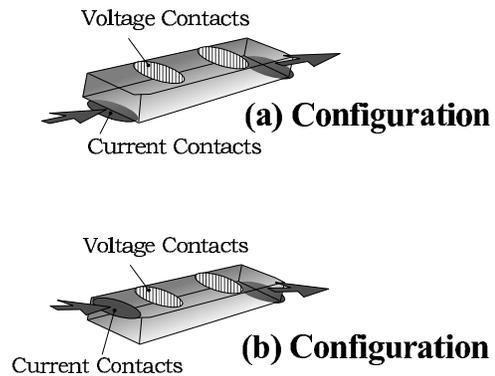}
\vspace{5mm}
\caption{Contact configurations used to measure the resistance of
the samples.} \vspace{1mm} \label{configuration}
\end{center}
\end{figure}

We have measured over 8 different parallelepiped twinned crystals of typical
dimensions 0.1x0.1x0.6 mm$^3$ under high pressure, which have given clearly
reproducible results. Depending on crystal, the electrical current was applied
parallel $(a)$ or non-parallel $(b)$ to its principal axis. These different
contact configurations are represented in Fig.~\ref{configuration}.  They were
initially chosen for the high pressure cell setup convenience. Due to the
large contact size respect to the small dimensions of the crystals, it was
meaningless to modify our contact setup in order to apply the Montgomery method
\cite{Montgomery71}  for a satisfactory determination of the anisotropic
resistivities. Although the obtained estimated resistivity values are within
the range of those previously reported\cite{Lakkis76,Schlenker80,Bartho69},
they only correspond to effective values, computed considering the conducting
area perpendicular to the principal axis direction, for the $(a)$ and $(b)$
configurations. From the structural point of view, according to a previous
work\cite{Bartho69}, Ti$_4$O$_7$ crystals have their longest dimension along
the $B$ crystallographic axis and show a metallic like conductivity at room
temperature. According to this, the $(a)$ contact configuration should be
testing essentially the conduction along the $B$ crystal direction, while the
$(b)$ configuration must include conductivity contributions of other
crystallographic directions. Nevertheless, by simulating an anisotropic
resistance network, we also checked that effectively the particular
resistivity dependencies with temperature obtained were not just a consequence
of the variation of the anisotropic resistivity ratio with temperature. Thus,
our approach, albeit certainly inexact, still remains as an effective way of
determining the general temperature behavior of the resistivity of these small
and anisotropic samples under high pressure.

\section{RESULTS}

We have analyzed the X-ray diffraction powder patterns  in order
to study the structure of the samples. The refined cell parameters
and angles of the triclinic unit cell (space group A-1) agree well
with literature values \cite{SAndersson59}. Lattice parameters
were determined for small single crystals by least-squares
refinement of the 2$\theta$ values of 25 very well-centered strong
reflections between 20 and 45 degrees. Cell parameters and angles
of the unit cell obtained for all crystals were analogous with
those expected for Ti$_4$O$_7$.

Samples measured using the $(a)$ contact configuration show a
metallic behavior at room temperature and a resistivity at ambient
pressure of $\simeq$ 2 m$\Omega$cm, while using the $(b)$ contact
configurations they display a semiconducting-like dependence and a
resistivity of $\simeq$ 10 m$\Omega$cm. They develop the
bipolaronic transitions,  with a 77 K to 300 K resistivity ratio
at ambient pressure of $\simeq$ 10$^8$ and $\simeq$ 10$^6$
depending on the contact configuration. These different values and
dependencies for the resistivity of Ti$_4$O$_7$ can be observed
comparing the data of previous papers \cite{Lakkis76,Schlenker80},
and have been interpreted as an intrinsic electrical transport
property related to the anisotropic conduction of this material
\cite{Inglis83}.

\begin{figure} [b]
\includegraphics[width=3.1in]{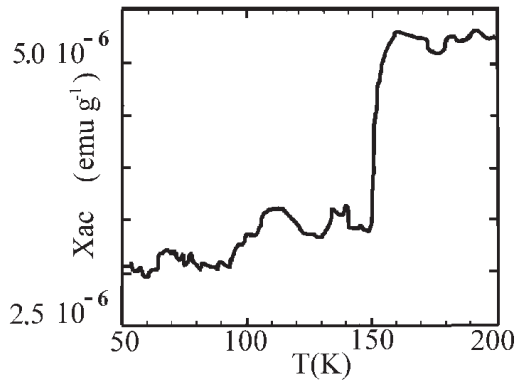}
\caption{ac susceptibility of Ti$_4$O$_7$ as a function of
temperature at ambient pressure. A sudden reduction of the Pauli
susceptibility can be seen for temperatures bellow $T_{bf}$
$\simeq$ 150 K.} \vspace{1.5mm} \label{Xac}
\end{figure}

Considering our experimental sensitivity of 10$^{-8}$ emu for the ac
susceptibility, various small Ti$_4$O$_7$ crystals were assembled together in
order to detect their small signal change when they undergo the bipolaron
formation  transition at $T_{bf}$. Despite the poor signal to noise ratio
obtained, the measured ac susceptibility, presented in Fig.~\ref{Xac}, shows
effectively a sudden reduction at $T_{bf} \sim$ 150 K and a small value for
lower temperatures, in accordance with previous results \cite{Lakkis76}.

\begin{figure} [t]
\includegraphics[width=3.1in]{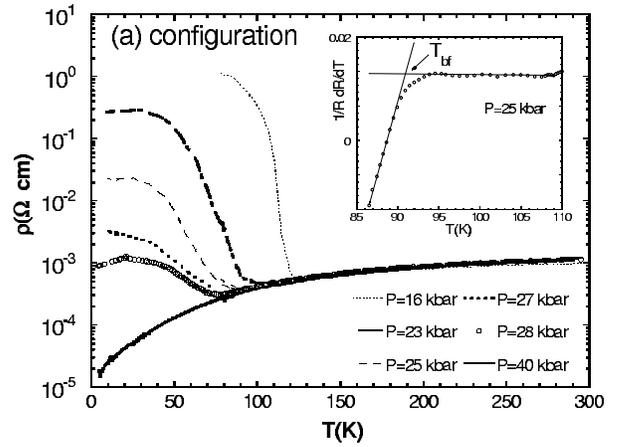}
\vspace{-0 mm}
\caption{Resistivity as a function of temperature of Ti$_4$O$_7$ in the $(a)$
configuration for different pressures (cooling). In the inset the criteria to
define $T_{bf}$ is presented. In this pressure range, only the transition at
$T_{bf}$ can be observed as the temperature of the bipolaron localized phase
($T_{bl}$) was abruptly reduced by pressure.} \vspace{-5mm} \label{raconf}
\end{figure}

\begin{figure} [hb]
\includegraphics[width=3.5in,height=2.6in]{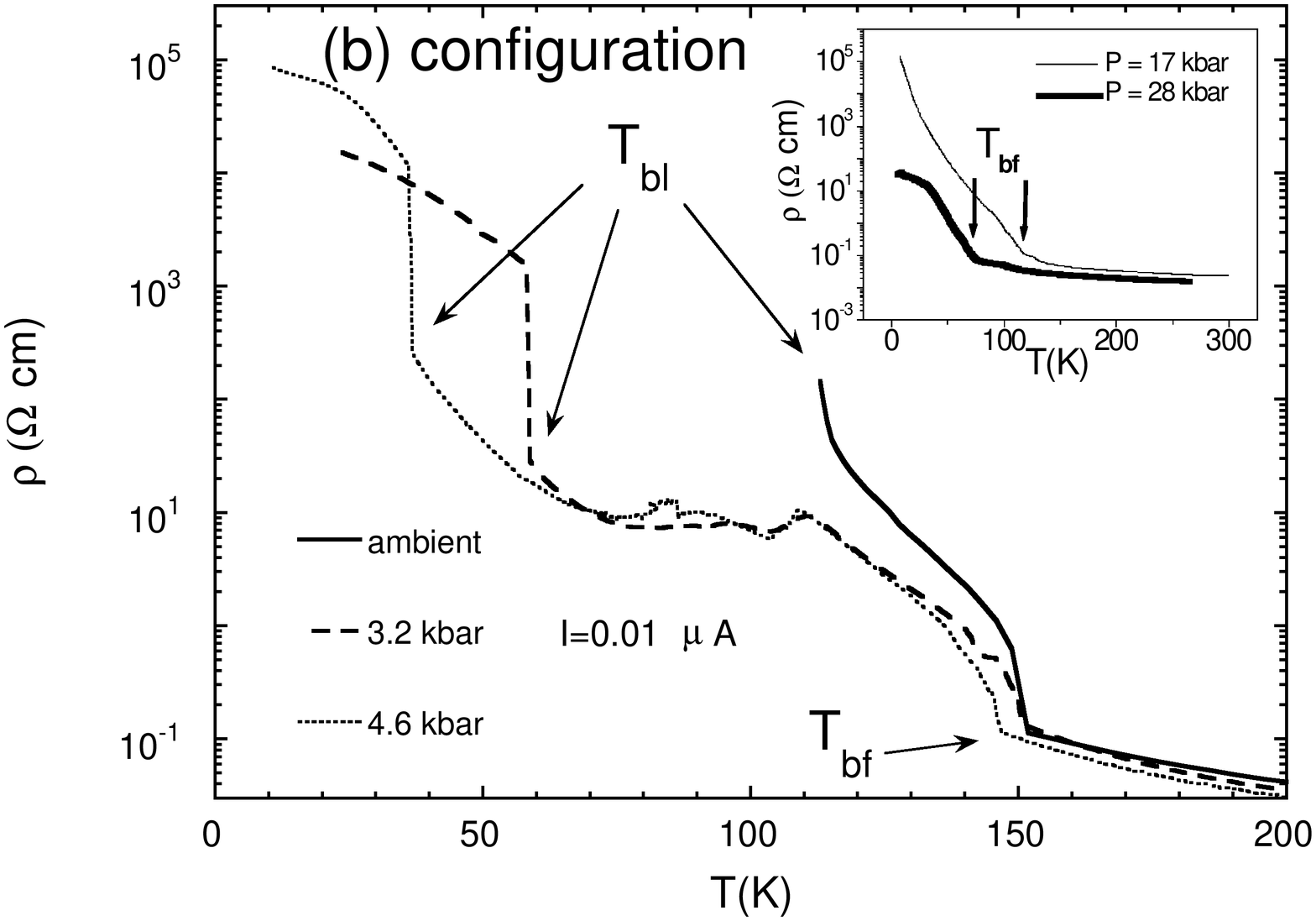}
\vspace{-5mm}
\caption{Resistivity as a function of temperature of Ti$_4$O$_7$ in the $(b)$
configuration for different pressures (cooling). The inset displays the
resistivity behavior for a higher pressure regime, where the bipolaron
localization transition at $T_{bl}$ is no longer present down to our minimum
experimental temperatures.} \vspace{0mm} \label{rbconf}
\end{figure}

In Fig.~\ref{raconf} and Fig.~\ref{rbconf} we can see $\rho (T)$  at different
applied pressures ($P$) for the $(a)$ and $(b)$ configurations, respectively.
Both transition temperatures ($T_{bf}$ and $T_{bl}$) decrease with increasing
pressure, although the $T_{bl}$ transition can be seen only in the hydrostatic
pressure setup as it is steeply decreased by a small pressure ($T_{bl}
\rightarrow 0$ for $P \simeq$ 4 kbar).

The localization transition temperature ($T_{bl}$) shows an hysteretical
behavior (not shown for clarity) which is consistent with the one previously
reported\cite{Lakkis76,Schlenker79}. A remanence of the metallic state can be
noticed in the low temperature range of the data of Fig.~\ref{rbconf} as a
positive d$\rho$/d$T$.

\begin{figure} [htbp]
\begin{center}
\includegraphics[width=3.8in]{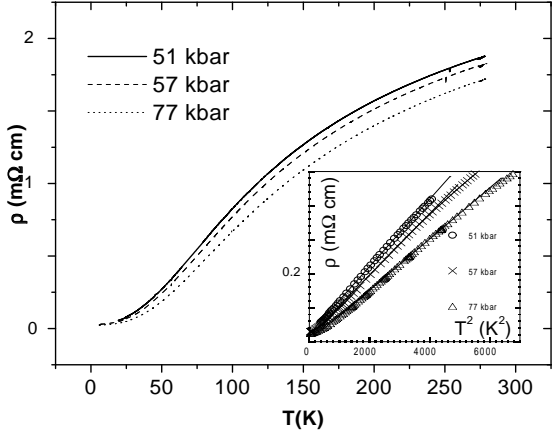}
\vspace{-5mm}
\caption{Resistivity for the high pressure regime (P$\geq$ 40
kbar), in the $(a)$ contact configuration. The inset shows that
resistivity can be represented, approximately by a $T^2$ law at
low temperatures.} \vspace{-5mm} \label{x3met}
\end{center}
\end{figure}

Now, if we focus our attention on the $(a)$ configuration results (shown in
Fig.~\ref{x3met}), we observe a poor metallic conduction for pressures $P
\geq$ 38 kbar ($\sim$ 2 m$\Omega$cm at room temperature), similar to those
observed for other transition-metal oxides \cite{McWhan69,Mott97}. The high
temperature behavior is sublinear while the low temperature resistivity shows
some signs of interacting carriers effects, as can be noticed in the inset of
Fig. \ref{x3met} as an approximate $T^2$ dependence of the resistivity.
Evidences of a superconducting transition were not observed down to 1.2 K over
this pressure range. For temperatures lower than $\sim$ 4 K a residual
resistivity $\rho_0 \sim$ 20-60 $\mu \Omega$ cm is obtained, which is reduced
with increasing pressure.

\begin{center}
\begin{figure}
\includegraphics[width=3.8in]{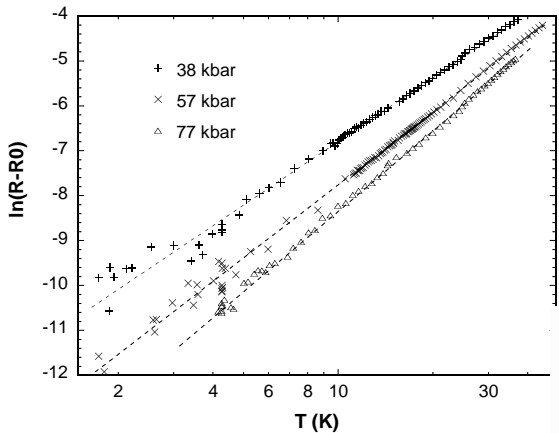}
\vspace{-0mm}
\caption{The logarithm of $\rho$ - $\rho_0$ is plotted vs T for different
pressures. The dashed lines were calculated according to the expression  $\rho
= \rho_0 + A T^N $.} \vspace{5mm} \label{qcp1}
\end{figure}
\end{center}

\begin{figure} [htbp]
\includegraphics[width=3.5in]{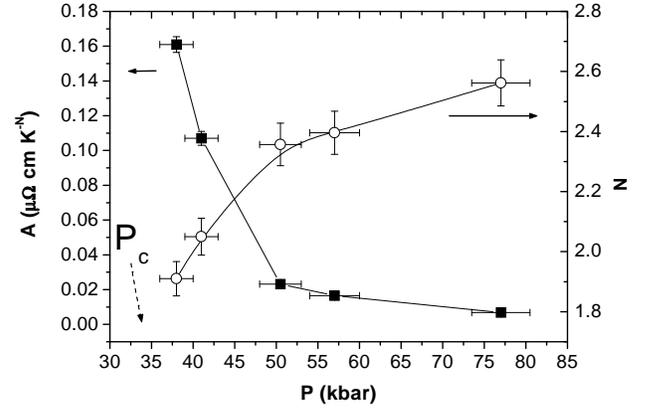}
\vspace{-2mm}
\caption{The $A$ and $N$ parameters determined by a fit of the
data shown in Fig. \ref{qcp1} to $\rho = \rho_0 + A T^N $ as a
function of pressure for 10 K $\leq$ $T$ $\leq$ 40 K.}
\vspace{5mm} \label{qcp2}
\end{figure}

Taking into account the possible proximity of a quantum critical point
(QCP)\cite{Sondhi97}, induced by the fact that $T_{bf}$, which defines a
paramagnetic metallic to non-magnetic insulator transition, probably tends to
0 for a critical pressure $P_c$ in the 34-37 kbar range, we evaluated the
coefficient $A$ and the exponent $N$ of the resistivity expression $\rho =
\rho_0 + A T^N $ as a function of pressure. For each pressure, a constant $N$
value is obtained for temperatures in the range 2 K $\leq T \leq$ 40 K. Results
are displayed in Fig. \ref{qcp1} and in Fig. \ref{qcp2}.

For $P$ $\rightarrow$ 34-37 kbar a divergence of the $A$ parameter
and the exponent $N$ $<$ 2 are observed. With increasing pressure,
the $N$ value increases, probably related to the relative increase
of the phonon contribution to the electronic scattering determined
by the $A$ coefficient diminution over one order of magnitude.

\begin{figure} [b]
\includegraphics[width=3.5in]{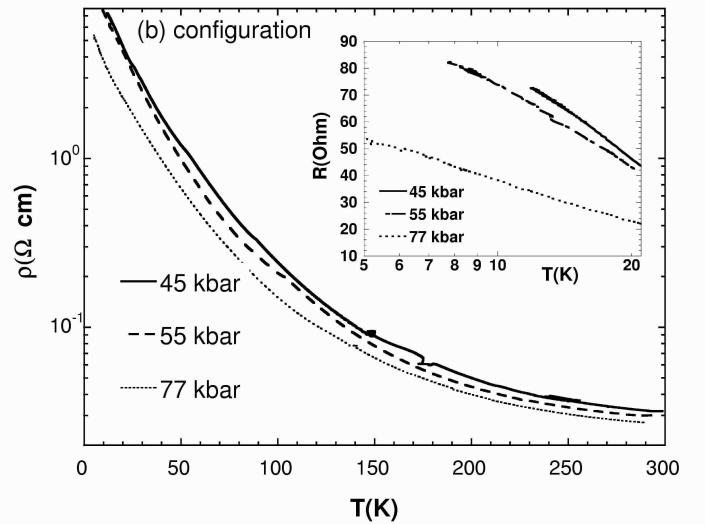}
\vspace{-0mm}
\caption{Temperature dependence of the resistivity at different pressures in
the $(b)$ contact configuration for a pressure regime where the transitions
are no longer seen down to our minimum experimental temperatures. The inset
shows a logarithmic behavior at low temperatures.} \vspace{-0mm} \label{logT}
\end{figure}

In Fig.~\ref{logT} we show our results using the $(b)$ contact configuration.
A very poor conduction is observed (10-100 m$\Omega$ cm, probably depending on
sample's twins and on current density, as it will be shown later) with a low
$P$ dependent behavior for pressures up to 230 kbar. Below 40 K a Ln(1/T)
dependence is observed, with a saturation regime at low temperatures. Both the
saturation resistance and the logarithmic slope decrease with increasing
pressure. A non-ohmic behavior was also detected in this region (see
Fig.~\ref{rvsIvsH}), but these studies will be published elsewhere.

Although our crystals are twinned, their electrical transport
characteristics were strictly determined by the chosen current
configuration setup ($(a)$ or $(b)$). As a consequence, this
indicates a particular correlation in the growing directions of
twins in these crystals. Considering this, the resistivity shown
for the $(a)$ configuration should be essentially related to a
mean value of the metallic resistivity tensor in the $B$ axis
direction, while the semiconducting-like resistivity related to
the $(b)$ configuration should be dominated by the resistivity
tensor in the $C$ crystallographic direction, where the shear
planes must act as a source of additional scattering.

\section{DISCUSSION}

\begin{figure} [ht]
\includegraphics[width=3.7in]{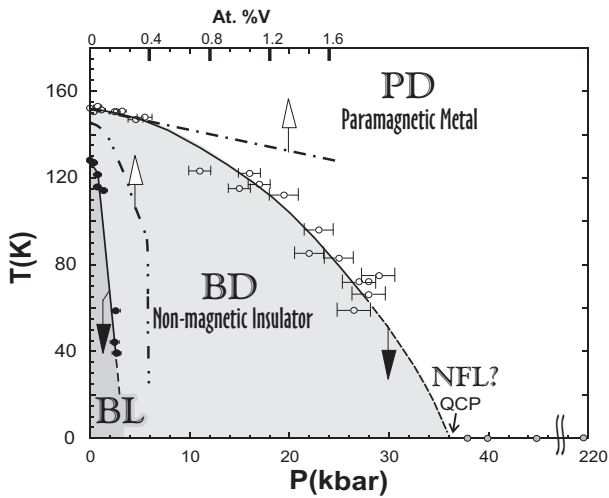}
\vspace{5 mm} \caption{The phase diagram shows the pressure dependence of
($\circ$) the bipolaron formation ($T_{bf}$) and ($\bullet$) localization
($T_{bl}$) transition temperatures. For comparison, the dot-dashed curves show
both transition temperatures for the V-doped samples. BL: bipolaron
localization. BD: bipolaron diffusion. PD: polaron diffusion.
 NFL: non-Fermi liquid. QCP: quantum critical point.}
\label{FasSchlC.eps} \vspace{-2 mm}
\end{figure}

Pressure modifies bipolaron ordering (localization) and formation
transition temperatures ($T_{bl}$ and $T_{bf}$) in a similar way
as V incorporation does in the (Ti$_{1-x}$V$_x$)$_4$O$_7$
compound\cite{Schlenker80}. A phase diagram, shown in Fig.
\ref{FasSchlC.eps}, was drawn including the electrical transport
characteristics observed for the $(a)$ and $(b)$ contact
arrangement, where the $T_{bl}$ transition was measured when
cooling the sample. For comparison, results of the V-doped samples
are also plotted but, in this case, the $T_{bl}$ transition was
determined by heating the sample. To do so, an arbitrarily
empirical scaling was established between the V content and the
applied pressure, as was determined for other systems
\cite{McWhan73}, forcing the V linear dependence of the bipolaron
formation temperature, $T_{bf}$(\%V), to match  the linear
$T_{bf}$($P$) dependency for low pressures ($P \leq$ 5 kbar).

\begin{figure} [h]
\includegraphics[width=3.6in]{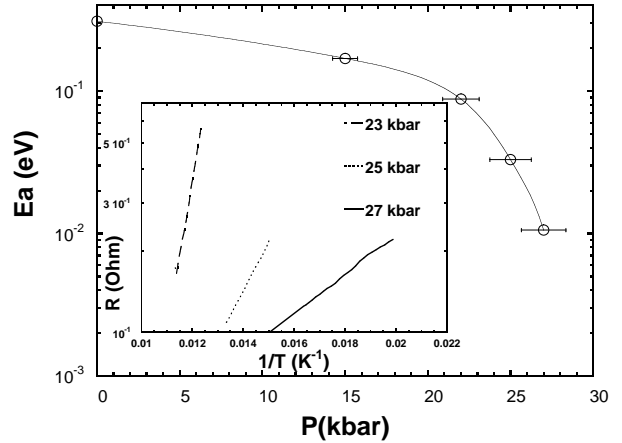}
\vspace{-0 mm}
\caption{Pressure dependence of the activation energy $E_a$ for the
resistivity of the ITP. The inset shows an Arhenius plot for the resistivity
at different pressures.} \vspace{-5mm} \label{EavsP}
\end{figure}

Although the similarity of the results is probably based on the fact that both
V incorporation and pressure decrease the Ti-Ti distance, generating  a local
chemical compression of the structure, there are some clear differences. The
effect of pressure on band structure and lattice modes seems to be more
significant than rigid band filling and impurities centers for the appearance
and ordering of bipolarons: The decrease and suppression of $T_{bl}$ is more
pronounced by applying an external pressure than by V incorporation. This also
states for $T_{bf}$ in the case of pressures $P >$ 5 kbar. Besides, when
$T_{bl}$ vanishes by the V-generated structural distortions, $T_{bf}$ is
smoothly reduced to $\sim$ 145 K. Contrary to this, pressure decreases
$T_{bf}$ down to $\sim$ 60 K, where it become difficult to distinguish the
associated resistive change. This is due to the fact that the activation
energy, $E_a$, that characterizes the electrical conduction of this
intermediate phase, is steeply reduced by pressure, as it can be seen in
Fig.~\ref{EavsP}.

As $E_a$ is expected to be the hopping energy of the bipolarons, which is
assumed to be half their binding energy \cite{Chakra76,Schlenker85}, this
pressure-induced sheer drop of $E_a$ is indicating an excess of conduction
probably related to hopping of single polarons. Within this picture, pressure
reduces the binding energy of bipolarons yielding to an electrical conduction
mediated by bipolarons and by single thermally excited polarons. In this
context, the metallic behavior observed in Fig.~\ref{rbconf} at low
temperatures can be associated with the coexistence of the high temperature
paramagnetic metallic phase and the low temperature bipolaronic insulator. For
small pressures, this phase separation can be related to the existence of
small metallic regions where the bipolaronic transition is not favored,
probably as a consequence of local structural distortions which inhibit the
formation of bipolarons. As pressure increases, the decrease of $E_a$ no
longer favors the binding of bipolarons, leading to a an increasing metallic
conduction.

\subsection{$(a)$ Contact configuration}

The approximated $T^2$ dependence of the resistivity in the $(a)$
contact configuration for pressures P$\geq$40 kbar (see the inset
of Fig. \ref{x3met}) reveals a large $T^2$ term which is
consistent with a picture of a highly correlated electron liquid.
A rough estimation of the $\gamma$ coefficient of the electronic
specific heat can be done considering the empirical relationship
between the $A$ and the $\gamma$ parameter established by Kadowaki
and Woods \cite{Kadowaki86}. A $\gamma \simeq$ 100 mJ/(mol K$^2$)
($P$=40 kbar, $N$=2) points out the magnitude of the
electron-electron interaction. The strong enhancement of the $A$
quantity would then be correlated with the approach to the
metal-insulator transition developed at $T_{bf}$. For the
temperature range considered, electron-phonon contributions can
not be minimized and the $N$ parameter can be a mean value fixed
by the weighted contributions of the electron-phonon  and the
electron-electron scatterings. The increase of the $N$ value with
increasing pressure can then be assigned to the reduction of the
electron-electron contribution suggested by the diminution of the
$A$ parameter.

Nevertheless more detailed measurements  are needed near the
possible critical pressure $P_c$ to determine if the divergence of
the $A$ parameter and, more important, the tendency to have an $N
<2$ are evidences that are suggesting a non-Fermi liquid behavior
(NFL), instead of the proximity of a metal-insulator transition.
Then, this possible NFL behavior  would be probably related to the
vicinity of a QCP \cite{Moriya95} for a critical pressure $P_c$ in
the 34-37 kbar range.  The QCP can be associated to the fact that
$T_{bf}$($P_c$)=0 K, setting up a phase transition between a
non-magnetic (bipolaronic-insulator) and a Pauli paramagnetic
(polaronic-metal) material at T=0 K. A similar behavior was
observed in BaVS$_3$ under pressure \cite{Forro00}.

\subsection{$(b)$ Contact configuration}

An interpretation of our results obtained for the $(b)$ contact
configuration must consider  the logarithmic divergence when
decreasing temperature.

\begin{figure}
\includegraphics[width=3.8in]{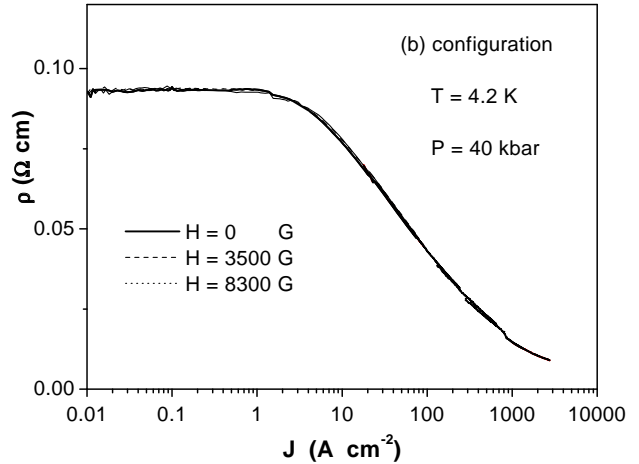}
\vspace{3mm}
\caption{Magnetic field sensitivity of the current dependent
resistivity at low temperatures in the $(b)$ contact
configuration. Within our experimental resolution, no appreciable
magnetic effects are observed.} \vspace{-5mm} \label{rvsIvsH}
\end{figure}

A logarithmic divergence in the resistivity can be found in dilute magnetic
alloys, or Kondo systems, where the conduction electrons are scattered with
magnetic impurities. Our Ti$_4$O$_7$ samples are clearly not a Kondo system,
as magnetic impurities are not supposed to be present, as was confirmed by its
low paramagnetic susceptibility at low temperatures, shown in the
Fig.~\ref{Xac}. Moreover, no appreciable magnetic field effects on the
resistivity were observed, as it is shown in Fig.~\ref{rvsIvsH}, contrary to
what we will expect for a Kondo system.

The logarithmic temperature dependence of the resistivity can also led us to
consider the framework used to describe electrical conduction in 2D systems
\cite{Lee85}, although, from the structural point of view, the oxygen
octahedra chains form a 3D interconnected network, so that a bidimensional
behavior can not be necessarily expected. In the case of a 2D conduction, a
magnetic field dependence of the resistivity should be obtained, which was not
observed experimentally as it is shown in Fig.~\ref{rvsIvsH}, although higher
magnetic field are necessary to obtain a conclusive result. Also, to our
knowledge, there are not bulk samples that show an electric conduction
characteristic of localized 2D systems.

As the presence of bipolaronic carriers in the high pressure phase
of Ti$_4$O$_7$ ($P \geq$ 40 kbar) can not be completely rejected,
we can also consider that the logarithmic divergence observed in
the $(b)$ contact configuration can be related to a diffusion of
bipolarons, scattered by random shallow potential wells
\cite{Alex97} in that particular direction. This disorder can be
generated by oxygen vacancies or by the intergrowth of another
members of the series Ti$_n$O$_{2n-1}$ which usually occurs along
the crystallographic $C$ axis\cite{LePage84}. In this case,
resistivity can be expressed at low temperatures (neglecting
localization effects) by the following equation:

\begin{equation}\label{eq:bips}
\rho(T) \simeq B Ln (\frac{E_0}{k_BT})
\end{equation}

\noindent where $B$ is a constant, $E_0$= $\pi^4 \hbar^2$ / 128
m$_B$ a$_{min}^2$, m$_B$ is the bipolaronic mass and a$_{min}$ is
the minimum size of the distribution of potential wells, which
should be near the interatomic spacing. The decrease of the
logarithmic slope with pressure (see the inset of Fig. \ref{logT})
can be interpreted, within this theoretical approach, as a
reduction of the product a$_{min}^2$ m$_B$, which is consistent
with a pressure-induced reduction of the interatomic spacing and
of the electron-phonon coupling which produces a decrease of the
bipolaron's mass.

Another possible interpretation of the observed logarithmic
divergence in the resistivity at low temperatures is related to
the electrical conduction in many metallic glasses
\cite{Cochrane75,Harris83}, where it is assumed that the
conduction electrons interact with atoms placed in a double
potential well, which can tunnel from one position to the other.
This tunneling between the two state configurations or two level
systems (TLS), which may be assisted by the electrons, generates a
new source of scattering for the conduction process. This effect
is usually not sensitive to the application of a magnetic field,
unlike for the Kondo alloys. Resistivity can be described by the
following empirical expression:

\begin{equation}\label{eq:rho}
\rho(T_{eff}) = C[1 - D Ln (T_{eff})]
\end{equation}

\noindent where

\begin{equation}\label{eq:TeffTLS}
T_{eff}^2 = T^2 + (\frac{\Delta}{ k_B})^2
\end{equation}

\noindent  $\Delta$  is the TLS energy separation and  $C$ and $D$
are suitable constants.

Although this model was initially  criticized
\cite{Black78,Black79} as it predict a contribution to the
resistivity one order of magnitude less than the experimental one,
further investigations opened the possibility of an incremental
effect of this contribution when many body problems are taken into
account \cite{Black81}.

The microscopic origin of TLS in Ti$_4$O$_7$ can be possibly
related to the different Ti site occupancy at low temperatures, or
to the dynamic equilibrium between two superstructures, revealed
in previous studies \cite{LePage84}. Also, it had been shown that
paired-electrons behave as TLS at low temperatures
\cite{Fox82,Phillips87}. There is no obvious theoretical
distinction between the motion of electrons accompanied by local
displacements of ions and the atomic motion proposed in the
tunneling model, so that the origin of the TLS can be
intrinsically related to bipolaronic diffusion.

\section{CONCLUSIONS}

An anisotropic electrical conduction with particular features was revealed
using different contact arrangements in Ti$_4$O$_7$ twinned crystals under
high pressure. Although a superconducting state was not achieved down to our
minimum experimental temperatures, a rich phase diagram was obtained, as shown
in Fig.~\ref{FasSchlC.eps}. Both formation and localization transition
temperatures are depressed increasing pressure.  The possible proximity of a
pressure-induced QCP was established in the metallic-like conduction contact
arrangement, as well as a conducting regime with clear signs of a highly
correlated electron liquid, for pressures over 40 kbar. A logarithmic
divergence in the resistivity with decreasing temperature was observed using
the non-parallel contact setup ($(b)$ contact configuration). In this latter
case, a valid explanation of the transport characteristics observed for the
whole temperature range studied seems to be closely related to an electrical
transport based on polarons (or excited bipolarons) scattered by TLS.

\section{acknowledgement}

This work was partially supported by CONICET of Argentina (PEI 146/98) and
Fundaci\'on Antorchas (A-13462/1). We also are indebted to J. Lorenzana, P.
Lee, J. Souletie and M. Rozenberg for fruitful discussions. We also acknowledge
technical assistance from C. Chiliotte, D. Gim\'enez, E. P\'erez Wodtke and D.
Rodr\'{\i}guez Melgarejo.


\begin{thebibliography}{10}

\bibitem{Chakra98}
B.~K. Chakraverty, J. Ranninger, and D. Feinberg, Phys. Rev. Lett.
{\bf 81},
  433  (1998).

\bibitem{Alex99}
A.~S. Alexandrov, Phys. Rev. Lett. {\bf 82},  2620  (1999).

\bibitem{Chakra99}
B.~K. Chakraverty, J. Ranninger, and D. Feinberg, Phys. Rev. Lett.
{\bf 82},
  2621  (1999).

\bibitem{Alex94}
A.~S. Alexandrov and N.~F. Mott, {\em High Temperature
Superconductors And
  Other Superfluids} (Taylor \& Francis, London, 1994).

\bibitem{Alex81}
A.~S. Alexandrov and J. Ranninger, Phys. Rev. B {\bf 23},  1796
(1981).

\bibitem{Iguchi88}
E. Iguchi, T. Yamamoto, and R.~J.~D. Tilley, J. Phys. Chem. Solids
{\bf 49},
  205  (1988).

\bibitem{Micnas90}
R. Micnas, J. Ranninger, and S. Robaszkiewicz, Rev. Mod. Phys.
{\bf 62},  113
  (1990).

\bibitem{Alex97}
A.~S. Alexandrov, Phys. Lett. A {\bf 236},  132  (1997).

\bibitem{Ando95}
Y. Ando {\it et~al.}, Phys. Rev. Lett. {\bf 75},  4662  (1995).

\bibitem{Ando96}
Y. Ando {\it et~al.}, Journal of Low Temperature Physics {\bf
105},  867
  (1996).

\bibitem{Schlenker74}
C. Schlenker, S. Lakkis, J.~M.~D. Coey, and M. Marezio, Phys. Rev.
Lett. {\bf
  23},  1318  (1974).

\bibitem{Chakra76}
B.~K. Chakraverty and C. Schlenker, J. Physique {\bf 37},  C4
(1976).

\bibitem{Lakkis76}
S. Lakkis, C. Schlenker, B.~K. Chakraverty, and R. Buder, Phys.
Rev. B {\bf
  14},  1429  (1976).

\bibitem{Schlenker79}
C. Schlenker, S. Ahmed, R. Buder, and M. Gourmala, J. Phys. C {\bf
12},  3503
  (1979).

\bibitem{Schlenker80}
C. Schlenker and M. Marezio, Phil. Mag. B {\bf 42},  453  (1980).

\bibitem{LePage84}
Y.~L. Page and M. Marezio, J. Solid State Chem. {\bf 53},  13
(1984).

\bibitem{Anderson75}
P.~W. Anderson, Phys. Rev. Lett. {\bf 34},  953  (1975).

\bibitem{Schlenker85}
C. Schlenker,  in {\em Bipolarons in transition metal oxides},
{\em Physics of
  disordered materials}, edited by D. Adler, H. Fritzsche, and S.~R. Ovschinsky
  (Plenum Press, NY, 1983), pp.\ 369--389.

\bibitem{Acha96}
C. Acha, M. N\'unez-Regueiro, M.~A.~A. Franco, and J. Souletie,
Czech. J. of
  Phys. {\bf 46},  2681  (1996).

\bibitem{SAndersson57}
S. Andersson, B. Collen, U. Kuylenstierna, and M. Magneli, Acta
Chem. Scand.
  {\bf 11},  1641  (1959).

\bibitem{Montgomery71}
H.~C. Montgomery, J. Appl. Phys {\bf 42},  2971  (1971).

\bibitem{Bartho69}
R.~F. Bartholomew and D.~R. Frankl, Phys. Rev. {\bf 187},  828
(1969).

\bibitem{SAndersson59}
S. Andersson {\it et~al.}, Acta Chem. Scand. {\bf 13},  989
(1959).

\bibitem{Inglis83}
A.~D. Inglis, Y.~L. Page, P. Strobel, and C.~M. Hurd, Solid State
Phys. {\bf
  16},  317  (1983).

\bibitem{McWhan69}
D.~B. McWhan and T.~M. Rice, Phys. Rev. Lett. {\bf 22},  887
(1969).

\bibitem{Mott97}
N.~F. Mott, {\em Metal-Insulator Transitions} (Taylor \& Francis,
London, 1997).

\bibitem{Sondhi97}
S.~L. Sondhi, S.~M. Girvin, J.~P. Carini, and D. Shahar, Rev. Mod.
Phys. {\bf
  69},  315  (1997).

\bibitem{McWhan73}
D.~B. McWhan {\it et~al.}, Phys. Rev. B {\bf 7},  1920  (1973).

\bibitem{Kadowaki86}
K. Kadowaki and S.~B. Woods, Solid State Commun {\bf 58},  507
(1986).

\bibitem{Moriya95}
T. Moriya and T. Takimoto, J. Phys. Soc. Jpn {\bf 64},  960
(1995).

\bibitem{Forro00}
L. Forr\'o {\it et~al.}, Phys. Rev. Lett {\bf 85},  1938  (2000).

\bibitem{Lee85}
P.~A. Lee and T.~V. Ramakrishnan, Rev. Mod. Phys. {\bf 57},  287
(1985).

\bibitem{Cochrane75}
R.~W. Cochrane, R. Harris, J.~O. Ström-Olson, and M.~J.
Zuckermann, Phys. Rev.
  Lett. {\bf 35},  676  (1975).

\bibitem{Harris83}
R. Harris and J.~O. Strom-Olsen,  in {\em Glassy Metals II},
No.~53 in {\em
  Topics in Applied Physics}, edited by H. Beck and H.~J. Guntherodt
  (Springer-Verlag, NY, 1983), pp.\ 325--342.

\bibitem{Black78}
J.~L. Black and B.~L. Gyorffy, Phys. Rev. Lett. {\bf 41},  1595
(1978).

\bibitem{Black79}
J.~L. Black, B.~L. Gyorffy, and J. Jackle, Philos. Mag. {\bf 40},
331  (1979).

\bibitem{Black81}
J.~L. Black,  in {\em Glassy Metals I}, No.~46 in {\em Topics in
Applied
  Physics}, edited by H. Beck and H.~J. Guntherodt (Springer-Verlag, NY, 1981),
  p.\ 167.

\bibitem{Fox82}
D.~L. Fox, B. Golding, and W.~H. Haemmerle, Phys. Rev. Lett. {\bf
49},  1356
  (1982).

\bibitem{Phillips87}
W.~A. Phillips, Reports on Prog. in Physics {\bf 50},  1657
(1987).

\end{thebibliography}

\newcommand{\noopsort}[1]{} \newcommand{\printfirst}[2]{#1}
  \newcommand{\singleletter}[1]{#1} \newcommand{\switchargs}[2]{#2#1}

\end{document}